\documentstyle[aps,epsf,multicol]{revtex}

\def\ket#1{\left | {#1} \right >}
\def\bra#1{\left < {#1} \right |}
\def\avg#1{\left < {#1} \right >}
\def\ie{{\it i.e.}}
\def\etal{{\it et. al. }}
\def\prb{Phys. Rev. B }
\def\prl{Phys. Rev. Lett. }

\begin{document}

\draft
\title{Study of transmission and reflection from a disordered lasing medium}
\author{Sandeep K. Joshi and  A. M. Jayannavar  }
\address{ Institute of Physics, Sachivalaya Marg, Bhubaneswar 751 005, India}

\maketitle

\begin{abstract}
A numerical study of the statistics of transmission ($t$) and reflection ($r$) of 
quasi-particles from a one-dimensional disordered lasing or amplifying medium 
is presented. The amplification is introduced via a uniform imaginary part
in the site energies in the disordered segment of the single-band tight binding
model. It is shown that $t$ is a non-self-averaging quantity. The cross-over
length scale above which the amplification suppresses the
transmittance is studied as a function of amplification strength.
A new cross-over length scale is introduced in the regime of strong disorder
and weak amplification.
The stationary distribution of the backscattered reflection 
coefficient is shown to differ qualitatively from the earlier analytical 
results obtained within the random phase approximation. 

\end{abstract}
\begin{multicols}{2}
\narrowtext
In recent years the study of wave propagation 
in an active random medium 
\cite{expt,Nkupp,abhi,pass,zhang,beenak,pusti,misir,freil1,john}
, in the presence of absorption or 
amplification, is being persued actively. The light propagation in an 
amplifying (lasing) medium has its implications for stimulated emission 
from random media. 
In the Schr\"odinger 
equation, to describe the absorption or amplification, one introduces the 
imaginary potentials. In that case the Hamiltonian becomes non-Hermitian and 
thus the particle number is not conserved. 
It should be noted that in quenched random systems  with imaginary potentials
the temporal 
coherence of the wave is preserved in spite of amplification or absorption 
which causes a particle non-conserving scattering process.

Some new results have been obtained in this area. 
In a scattering problem, the 
particle experiences a mismatch from the real valued potential to the imaginary valued
potential at the interface between the free region and the absorbing (or amplifying)
medium, and hence it tries to avoid this region by enhanced back reflection. Thus
a dual role is played by imaginary potentials as an absorber (or amplifier) and as a
reflector.  
When the strength of the imaginary potential
is increased beyond certain limit, both absorber and amplifying scatterer act 
as reflectors. Thus the reflection coefficient exhibits 
non-monotonic behavior as a function of the absorption (amplification)
strength. Using the duality relations it has been shown that 
amplification suppresses the transmittance in the large length (L) limit just
as much as absorption does \cite{pass}.  
There exists a crossover length scale $L_c$
below which the amplification enhances the transmission and above which
the amplification reduces the transmission which, in fact, vanishes
exponentially in the $L \rightarrow \infty$ limit. In contrast, reflectance
saturates to a finite value. Moreover,
absorption and amplification of same strength (\ie, differing only 
in the sign of the imaginary part) will induce same localization length \cite{beenak}. 

In an amplifying medium even though the transmittance ($t$) decreases exponentially
with the length $L$ in the large $L$ limit, the average $\avg{t}$ is shown \cite{pusti}
to be infinite due to the less probable resonant realizations corresponding to
the non Gaussian tail of the distribution of $ln~t$. This result is based on
the analysis using random phase approximation (RPA). Using duality argument Paasschens
\etal show that non-Gaussian tails in the distribution of $lnt$ contain negligible weight \cite{pass}.
Thus one might expect finite value for $\avg{t}$ in the asymptotic
limit. It should be noted that even in the ordered periodic system all the states
are resonant states and still the transmittance decreases exponentially for
all the states in the large $L$ limit. The above simple case may indicate that 
in the asymptotic limit $\avg{t}$ is indeed finite. One of our objectives in 
this paper is to study the behavior of the transmission 
probability as a function of length in the presence of coherent amplification. 
We show that the transmission coefficient is a 
non-self-averaging quantity. In the large length limit we do not find any resonant
realization, which can give an enhanced transmission. We also study the behavior 
the of cross-over length $L_c$
as a function of amplification strength. 
We have analysed the behavior of logarithm of the transmittance which
will have a maximum value $\avg{lnt}_{max}$ at $L_c$, 
and its dependence on the amplification strength. 
For a given strength of 
amplification there exists a critical strength of disorder below which the
average transmittance is always less than unity at all length scales and
decreases monotonically. In this regime $L_c$ and $\avg{lnt}_{max}$ loose their 
physical significance. In this regime we show that there exists a new cross-over
length scale $\xi_c$ which diverges as the amplification strength is
reduced to zero for a given strength of the disorder.

In the work by Pradhan and Kumar \cite{Nkupp}, the analytical expression for the stationary
distribution $P_{s}(r)$ of a coherently backscattered reflection coefficient $(r)$
is obtained in the presence of amplification in a  RPA
 using the method
of invariant imbedding. 
The expression for $P_{s(r)}$ is given by
\begin{eqnarray}
P_s(r) & = & \frac{|D|exp(-\frac{|D|}{r-1})}{(r-1)^2}~~~for~~ r \geq 1 \label{prad} \\
       & = &  0~~~~~~~~~~~~~~~~~                     ~~~for~~ r < 1 \nonumber
\end{eqnarray}
where $D$ is proportional to $\eta / W$, $\eta$ and $W$ being the 
strength of amplifying potential and disorder respectively. One can readily
notice from Eqn. (\ref{prad}) that $P_s(r)$ does not tend to $\delta (r-1)$ in
the large $\eta$ limit. In this limit, as mentioned earlier, an amplifying 
scatterer acts as a reflector. 
The validity of
above expression is limited to 
small disorder and amplification strength. 

We consider a quasi-particle moving on a lattice. The appropriate Hamiltonian in
a tight-binding one-band model can be written as 

\begin{equation}
H~=~\sum \epsilon_n^\prime \ket{n}\bra{n} + V(\ket{n}\bra{n+1}~+~\ket{n}\bra{n-1} ).
\end{equation}

$V$ is the off-diagonal matrix element connecting nearest neighbors separated
by a lattice spacing $a$ (taken to be unity throughout) and $\ket{n}$ is the
non-degenerate Wannier orbital associated with site $n$, where $\epsilon_n^\prime
=\epsilon_n-i\eta$ is the site energy. The real part of the site energy
$\epsilon_n$ being random represents static disorder and $\epsilon_n$ at
different sites are assumed to be uncorrelated random variables distributed
uniformly ($P(\epsilon_n)=1/W$) over the range $-W/2$ to $W/2$. 
We have taken imaginary part of the
site energy $\eta$ to be spatially uniform 
positive variable signifying amplification. Since all
the relevant energies can be scaled by $V$, we can set $V$ to unity. The lasing
medium consisting of $N$ sites ($n=1$ to $N$) is embedded in a perfect infinite
lattice with all site energies taken to be zero.
To calculate the transmission and reflection coefficients we use the well known
transfer-matrix method , and the details are
described in Ref. \cite{abhi} .

In our studies we have set the energy of the incident particle 
at $E=0$, \ie, at a midband energy. Any other value for the incident
energy does not affect the physics of the problem. In calculating average
values in all cases we have taken 10,000 realizations of random site energies
($\epsilon_n$). The strength of the disorder and the amplification are scaled
with respect to $V$, \ie, $W~(\equiv W/V)$ and $\eta~(\equiv \eta/V)$.
The length $L$ = $L/a$. 

Depending on the parameters $\eta$, $W$ and $L$ the transmission coefficient
can be very large (of the order of $10^{12}$ or more). Hence, we first
consider behavior of $\avg{lnt}$ instead of $\avg{t}$. The angular
brackets denote the ensemble average. 
The localization length $\xi$ can be computed from the
behaviour of transmittance in the large length limit. We denote
the localization length $l_a$for an ordered medium (W=0)in the presence
of uniform amplification.  
The localization length \cite{econ} for
a disordered passive medium ($\eta=0$) is given by elastic back scattering
length $l=48V^2/W^2$ at the center of the band ($E=0$). We have verified
that the localization length in the presence of both disorder and
amplification, \cite{zhang} $\xi$ is related to $l$ and $l_a$ (for $\eta/V < 1$
and $W/V < 1$) as $\xi=ll_a/(l+l_a)$.

\begin{figure}
\protect\centerline{\epsfxsize=3in \epsfysize=3in \epsfbox{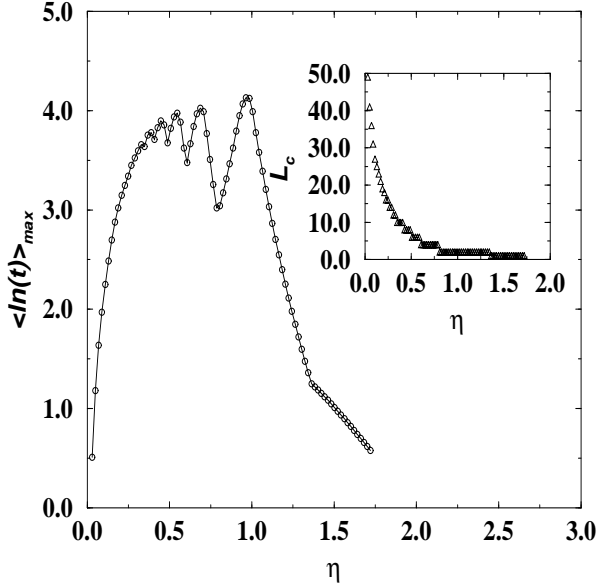}}
\vspace{0.5in}
\caption{The variation of $\avg{lnt}_{max}$ with amplification strength $\eta$ 
for $W=1.0$. Inset shows the variation of $L_c$ with $\eta$ for $W=1.0$.}
\label{mteta}
\end{figure}

In Fig. \ref{mteta} we have plotted $\avg{lnt}_{max}$ against $\eta$ for a 
fixed value of $W=1.0$ and the inset shows variation of $L_c$ with $\eta$
for $W=1.0$. Initially $\avg{lnt}_{max}$ increases with $\eta$ and after 
exhibiting a maxima it decays to zero for large $\eta$. This arises from
the fact that the lasing medium acts as a reflector for large $\eta$ as
discussed in the introduction. Near the maximum, in a finite regime of $\eta$,
$\avg{lnt}_{max}$ exhibits several oscillations. In this region sample to 
sample fluctuations of $lnt$ are very large. Thus average over $10,000$ 
realizations may not represent the true ensemble averaged quantity. From
the curve fitting of our numerical data for $L_c$, we find that $L_c$ does not follow 
a power law, ($1/\sqrt{\eta}$), in the full parameter regime \cite{zhang}.

\begin{figure}
\protect\centerline{\epsfxsize=3in \epsfysize=3in \epsfbox{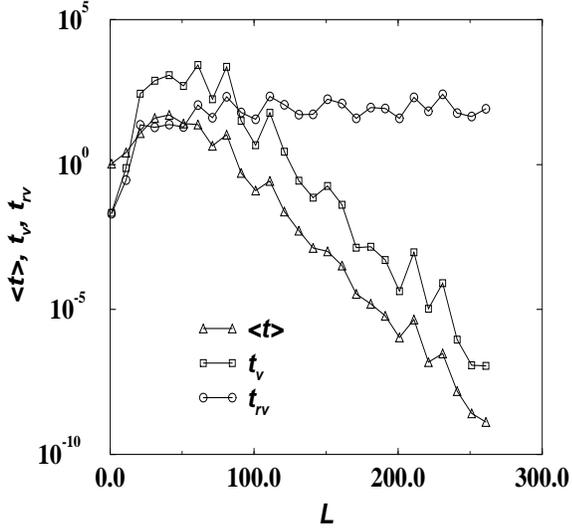}}
\vspace{0.5in}
\caption{The plot of $\avg{t}$, root-mean-squared variance ($t_v$) and root-mean-squared
relative variance ($t_{rv}$) as a function of length $L$ for $\eta=0.1$ and $W=1.0$.}
\label{varT}
\end{figure}

To study the nature of fluctuations in the transmission coefficient, in Fig. 
\ref{varT} we have plotted, on log-scale, $\avg{t}$, root-mean-squared variance $t_v =
\sqrt{\avg{t^2}-\avg{t}^2}$ and root-mean-squared relative variance (or
fluctuation) $t_{rv}=\sqrt{\avg{t^2}-\avg{t}^2}/\avg{t}$ as a function of
$L$ for $\eta=0.1$ and $W=1.0$. For these parameters $l \approx 48$,$l_a = 10$,
$\xi \approx 8$ and $L_c \approx 30$. We notice that both $\avg{t}$ and $t_v$
exhibit maxima and decrease as we increase the length further. Except in the small length
limit, variance is larger than the mean value. The relative variance is larger than 
one for $L>10$ and remains large even in the large length limit. The $t_{rv}$ fluctuates
between values $50$ to $300$ in the large length ($L>10$) regime, indicating
clearly the non-self-averaging nature of the transmittance. This implies that
the transmission over the ensemble of macroscopically 
identical samples dominates
the ensemble average.  
In such a situation one has to consider the full probability distribution $P(t)$
of $t$ to describe the system behavior.

\begin{figure}
\protect\centerline{\epsfxsize=3in \epsfysize=3in \epsfbox{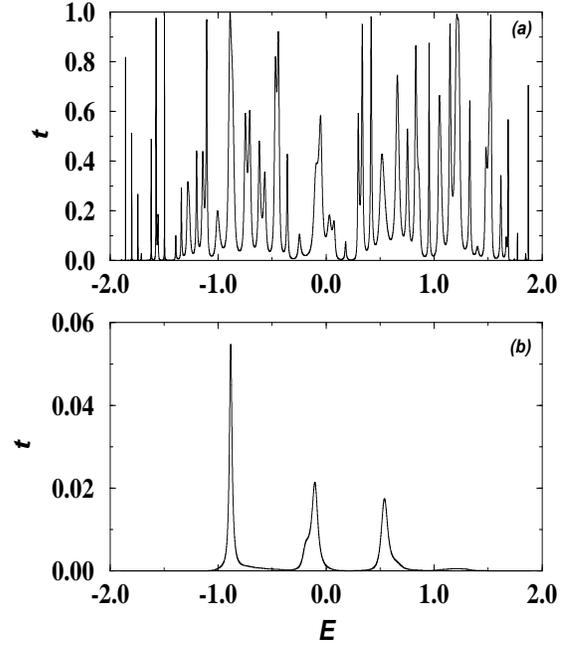}}
\vspace{0.5in}
\caption{Transmittance $t$ as function of incident energy $E$ for $W=1.0$,
$L=100$ and (a) $\eta=0$ and (b) $\eta=0.1$}
\label{TvsE}
\end{figure}

We would now like to understand 
whether there exist any resonant realizations in 
the large length limit for which the 
transmittance is very large. This study calls
for sample to sample fluctuations. It is well known  from the studies in passive
random media that the ensemble 
fluctuation and the fluctuations for a given sample
as a function of chemical potential or 
energy are expected to be related by some 
sort of ergodicity \cite{ergod}, \ie, the measured fluctuations as a function of
the control parameter are identical to 
the fluctuations observable by changing the
impurity configurations. 
In Fig. \ref{TvsE}(a) we have plotted $t$ versus incident 
energy $E$ (within the band from $-2$ to $+2$) 
for a given realization of random 
potential with $\eta=0$ and $L=100$. 
The Fig. \ref{TvsE}(b) shows the behavior of $t$ versus
$E$ for the same realization in the presence of amplification $\eta=0.1$ and $L=100$. 
>From the Fig. \ref{TvsE}(a) we observe that at several
values of energy the transmittance 
exhibits the resonant behavior in that $t=1$. 
>From Fig. \ref{TvsE}(b) we notice
that in the presence of amplification, transmittance at almost resonant realizations is 
negligibly small. Few peaks appear in the transmittance whose origin lies in the 
combined effect of disorder and amplification. However, we notice that the transmittance 
at these peaks is much smaller, where as one would 
have naively expected the transmittance to be much much larger than unity in the amplifying
medium. We have studied several realizations and found that none of them shows any 
resonant behavior where one can observe the large transmittance. The peak value of 
observed transmittance is of the order of unity or less. This study clearly indicates 
that $\avg{t}$ is indeed finite contrary 
to the earlier predictions based on RPA \cite{pusti}.

\begin{figure}
\protect\centerline{\epsfxsize=3in \epsfysize=3in \epsfbox{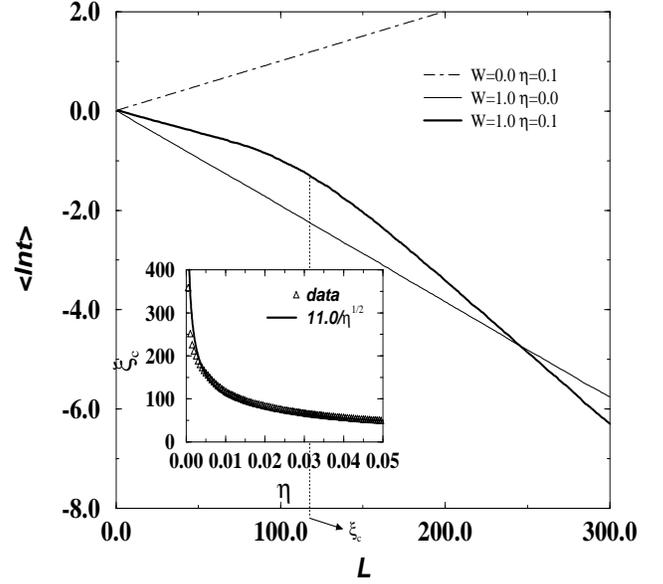}}
\vspace{0.5in}
\caption{Variation of $\avg{lnt}$ with $L$. The new length scale $\xi_c$ which
arises for $\eta \ll 1.0$ is shown by a vertical dotted line. The inset shows
the variation of $\xi_c$ with $\eta$ for $W=1.0$. The numerical fit shown by
the thick line indicates that $\xi_c$ scales as $\eta^{-1/2}$ in this regime.}
\label{ltl}
\end{figure}

So far our study was restricted to the parameter space of $W$ and $\eta$ for
which $L_c$ and hence $\avg{lnt}_{max}$ exist. In Fig. \ref{ltl} we have plotted
$\avg{lnt}$ against $L$ for ordered lasing medium ($W=0$,$\eta=0.01$), disordered
passive medium ($W=1.0$,$\eta=0$) and disordered active medium ($W=1.0$,$\eta=0.01$).
The present study is restricted to the parameter space of $\eta$ and $W$ such that
$\eta \ll 1.0$ and $W \geq 1.0$. We notice that for an ordered lasing medium, the
transmittance is larger than one.
We have taken our range of $L$ upto $300$. 
For a disordered active medium ($W=1.0$,$\eta=0.01$), we notice that the transmittance
is always less than one and monotonically decreasing. Initially, upto certain
length, the average transmittance is, however, larger than that in the disordered passive
medium ($W=1.0$,$\eta=0$). This arises due to the combination of lasing with disorder.
In the asymptotic regime transmittance of a lasing random medium falls below that in the
passive medium with same disorder strength. This follows from the enhanced localization
effect due to the presence of both disorder and amplification together, \ie, $\xi < l$.
It is clear from the figure that $\avg{lnt}$ does
not exhibit any maxima and hence the question of $L_c$ or $\avg{lnt}_{max}$ does
not arise. We notice, however, from the figure that for random active medium initially
$\avg{lnt}$ decreases with a well defined slope and in the large length limit
$\avg{lnt}$ decreases with a different slope (corresponding to localization
length $\xi$). Thus we can define a length scale $\xi_c$ (as indicated
in the figure) at which there is a cross-over from the initial slope to the 
asymptotic slope. In the inset of Fig. \ref{ltl} we have shown the dependence 
of $\xi_c$ on $\eta$. Numerical fit shows that $\xi_c$ scales
as $1/\sqrt{\eta}$, as we expect  $\xi_c \rightarrow \infty$ with $\eta 
\rightarrow 0$. As one decreases $\eta$, the absolute value of initial slope increases
and that of the asymptotic one decreases. Simultaneously, the cross-over length
$\xi_c$ increases. In the $\eta \rightarrow 0$ limit both initial as well
as asymptotic slopes become identical.

\begin{figure}
\protect\centerline{\epsfxsize=3in \epsfysize=3in \epsfbox{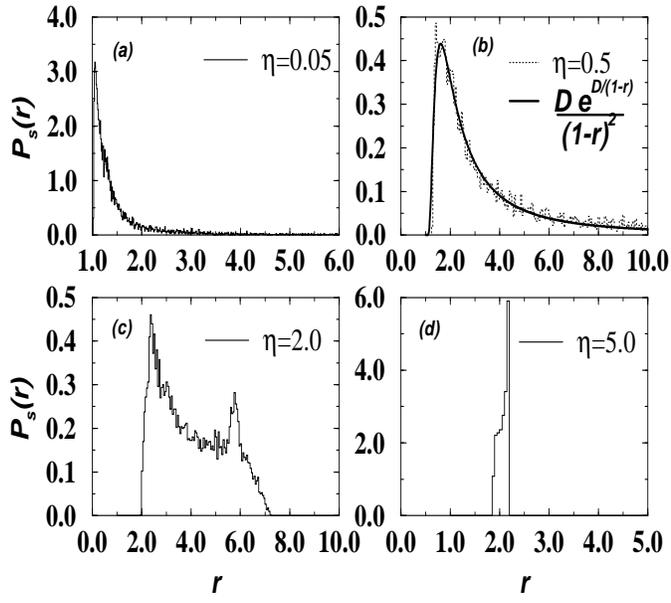}}
\vspace{0.5in}
\caption{Stationary distribution of reflection coefficient $P_s(r)$ for $W=5.0$ and 
various values of $\eta$. The numerical fit shown in Fig. \ref{psr}(b)
with a thick line has $D=1.235$}
\label{psr}
\end{figure}

In Fig. \ref{psr} we have plotted the stationary distribution $P_s(r)$ of
reflection coefficient $r$ for different values of $\eta$ (as shown in the
figure) and a fixed value of $W=5.0$. 
For small values
of $\eta=0.05$ the stationary distribution $P_s(r)$ has a single peak around
$r=r_{max}=1$. The peak ($r_{max}$) shifts to higher side as we increase $\eta$
(Fig. \ref{psr}(b)). The behavior of $P_s(r)$ for small $\eta$ is in qualitative
agreement with Eqn. (\ref{prad}). As we increase $\eta$ further $P_s(r)$ 
exhibit a double peaked structure (Fig. \ref{psr}(c)). At first the second
peak appears at higher value of $r$ at the expense of the distribution at 
the tail. As we increase $\eta$ the second peak becomes more prominent and 
shifts towards left, where as the height of the first peak decreases. The
distribution at the tail has a negligible weight (see Fig. \ref{psr}(c)). At
still higher values of $\eta$, the second peak approaches $r \approx 1$ whereas 
the first peak disappears. The now-obtained single peak distribution $P_s(r)$ in
the large $\eta$ limit tends to $\delta(r-1)$. In this limit the amplifying medium 
acts as a reflector and the disorder plays a sub-dominant role. The occurrence of
the double peak structure along with $P_s(r) \rightarrow \delta(r-1)$ in the large
$\eta$ limit cannot be explained even qualitatively from Eqn. (\ref{prad}).
This is due to the failure of RPA in this regime \cite{abhi}.

Our study for $\avg{lnr}_s$, obtained from $P_s(r)$ indicate that it 
initially increases with $\eta$ and exhibiting a maxima at $\eta_{max}$
decreases monotonically. The double peak in $P_s(r)$ appears for values
close to $\eta_{max}$. For the value $\eta > \eta_{max}$ the amplifying medium
acts dominantly as a reflector. In this regime  we have verified that
the stationary distribution of the phase $\theta$ of the
complex reflection amplitude is not uniform.
Physics of the double peak and overall shape of $P_s(r)$ shows similarity with the 
stationary distribution obtained in the case of absorption (for details we
refer to Ref. \cite{abhi}.

In conclusion
our numerical study on the statistics of transmission coefficient in 
random lasing medium indicates that in the asymptotic regime the 
transmission coefficient is a non-self-averaging quantity,however, with a well 
defined finite average value. 
In some parameter space transmittance initially increases with $\eta$
and falls off exponentially to zero in the asymptotic regime. In this
regime there is a well defined cross-over length $L_c$ at which the
transmittance is maximum, and it decreases monotonically with $\eta$.
In the parameter range where $\eta \ll 1$, in the
presence of disorder the average transmittance decreases monotonically
and has a magnitude less than unity. In this regime $L_c$ does not exist.
However, one can still define a new length scale $\xi_c$ which
scales as $1/\sqrt{\eta}$. Our study on the stationary distribution
of reflection coefficient $P_s(r)$ indicates that earlier analytical
studies fail, even qualitatively, to explain the observed behavior
in the large $\eta$ limit. 
Our study clearly brings out the dual role played by an amplifying medium,
as an amplifier as well as a reflector.

\end{multicols}
\end{document}